\documentclass[aps,pra,twocolumn, noshowpacs, floatfix]{revtex4}
\usepackage{amsfonts}
\usepackage{amssymb}
\usepackage{graphicx}
\usepackage{amsmath}
\usepackage[english]{babel}
\usepackage{color}
\usepackage{epstopdf}
\usepackage{footnote}
\begin{document}
	\title{Cooperative Suppression of Two-Photon Excitation by Kerr Nonlinearity and Cavity Loss in a Driven Holstein–Tavis–Cummings System}
\author{Zhiyao Qu$^{1}$, Kewei Sun$^{1}$\footnote{Electronic address:~\url{skw79724@hdu.edu.cn}}, Youhao Chang$^{2,3}$, Vladimir Al. Osipov$^{4}$, Yang Zhao$^{5}$, and Maxim F. Gelin$^{1}$\footnote{Electronic address:~\url{maxim@hdu.edu.cn}}}
\affiliation{$^{1}$\mbox{Department of Physics, Hangzhou Dianzi University, Hangzhou 310018, China}\\}
\affiliation{$^{2}$\mbox{School of Physics and Optoelectronic Engineering, Hainan University, Haikou 570228, China}\\}
\affiliation{$^{3}$\mbox{Center of Theoretical Physics, Hainan University, Haikou 570228, China}\\}
\affiliation{$^{4}$\mbox{Institute for Advanced Study in Mathematics, Harbin Institute of Technology, Harbin 150001, China}\\}
\affiliation{$^{5}$\mbox{School of Materials Science and Engineering, Nanyang Technological University, Singapore 639798, Singapore}\\}
\begin{abstract}

We apply the numerically accurate multiple Davydov Ansatz (multi-$\rm D_2$) method to investigate the nonequilibrium dynamics of an open molecular cavity system using an extended driven-dissipative Holstein-Tavis-Cummings  model incorporating Kerr nonlinearity. Our results reveal that the synergy between Kerr nonlinearity and exciton-phonon coupling induces an anharmonic level structure, imposing a stringent energy penalty that isolates the one-excitation manifold. This establishes a robust single-excitation-dominated regime that dynamically manifests photon blockade, characterized by suppression of the pure two-photon state and isolation of the one-excitation manifold. While strong coherent driving induces power broadening that overcomes the nonlinear barrier and opens multiphoton channels, cavity dissipation counteracts this process by acting as a highly selective manifold filter. By preferentially depleting fragile multi-photon states while sheltering single excitations within dissipation-free dark states, the  cavity mode  decay rate actively cooperates with Kerr nonlinearity to restore and reinforce the photon blockade effect. These findings elucidate the intricate interplay among nonlinearity, driving, and dissipation, offering a theoretical blueprint for engineering robust single-photon sources in open molecular cavity platforms.

\end{abstract}
\date{\today}
\maketitle

Cavity quantum electrodynamics (CQED) provides a fundamental framework for exploring the coherent interaction between quantized light fields and matter in spatially confined environments~\cite{b1}. Driven by recent advances in micro- and nanofabrication, cryogenic control, and precision quantum measurement techniques, modern CQED systems have become premier platforms for accessing the strong light-matter coupling regime~\cite{b2,b3,b4,b5,b6}. In this regime, the rapid exchange of energy between light and matter gives rise to a wealth of striking quantum phenomena, including vacuum Rabi splitting~\cite{b7,b8,b9,b10,b11,b12}, quantum entanglement~\cite{b13,b14}, quantum optical bistability~\cite{b15}, and  photon blockade~\cite{b16,b17,b18}.

The theoretical description of light-matter interaction traces back to the seminal Jaynes-Cummings (JC) model, which captures the coupling between a single-mode light field and a single two-level atom~\cite{b19}. This was subsequently generalized by Tavis and Cummings to the Tavis-Cummings (TC) model, describing multiple two-level systems interacting with a single-mode field, thereby laying the foundation for understanding collective coupling, superradiance, and polariton phenomena in many-body systems~\cite{b20,b21}. However, in realistic solid-state systems and molecular aggregates, degrees of freedom beyond electronic excitations, such as molecular vibrations, lattice phonons, and environmental fluctuations significantly affect energy-level structures, decoherence processes, and nonequilibrium dynamics~\cite{b22,b23,b24}. To address this, the Holstein-Tavis-Cummings (HTC) model incorporates Holstein-type exciton-phonon coupling into the TC framework~\cite{b25,b26}. By simultaneously including cavity photons, molecular excitons, and vibrational phonons, the HTC model has been widely employed to investigate polariton formation, energy transfer, vibration-assisted transport, and related nonequilibrium dynamics in organic microcavities~\cite{b27,b28,b29,b30}.

Despite its success, standard HTC studies predominantly focus on closed systems or linear regimes. In realistic molecular cavity platforms, however, the system is inherently open and subject to intrinsic photonic-mode nonlinearities, external coherent driving, and cavity dissipation~\cite{b31,b32,b33}. The interplay between many-body interactions (exciton-phonon coupling) and photonic nonlinearities in such driven-dissipative environments remains poorly understood. Specifically, how these combined effects renormalize the energy level structure and modify the nonequilibrium population transfer pathways is a critical open question.

A key nonlinear phenomenon in this context is photon blockade~\cite{b16,b17,b18}. It is a prototypical quantum optical effect manifested as photon antibunching, where multiphoton excitation is suppressed and photons are emitted predominantly one at a time~\cite{b16,b34}. This effect provides a crucial mechanism for realizing single-photon sources and on-chip quantum optical devices~\cite{b17,b35,b36,b37}. To engineer such anharmonic energy-level structures and control multiphoton excitation processes, Kerr nonlinearity serves as an essential mechanism~\cite{b38,b39,b40}. When combined with external coherent driving (which injects energy) and cavity dissipation (which induces energy losses), these elements collectively govern the complex nonequilibrium dynamics of the open molecular cavity system.

From a computational perspective, many-body quantum systems involving photonic, excitonic, and phononic degrees of freedom possess high-dimensional Hilbert spaces, making analytical solutions intractable. To study the real-time dynamics of such light-matter-phonon coupled systems, several advanced numerical approaches have been developed and widely applied. These include the multi-configuration time-dependent Hartree (MCTDH) method~\cite{b41,b42,b43,b44}, tensor-train and matrix product-state methods~\cite{b45,b46,b47,b48}, and the Davydov Ansatz (DA) formalism~\cite{b49,b50}. Among these, the DA and its multi-configurational extensions (mDA) offer a promising  balance of clear physical interpretation and high computational efficiency, having been successfully applied to Holstein molecular crystals, spin-boson models, energy transfer processes, and polaron dynamics~\cite{b51,b52,b53,b54}. In particular, the so-called multi-$\rm D_2$ Ansatz improves the flexibility and accuracy of the trial wave function through the superposition of multiple coherent states, enabling a rigorous description of ground-state properties and nonequilibrium dynamics in low-temperature quantum many-body systems~\cite{b55,b56}.

Motivated by the above  considerations, we construct a generalized HTC model by incorporating Kerr nonlinearity, external coherent driving, and cavity dissipation. Employing the multi-$\rm D_2$ Ansatz combined with the time-dependent variational principle, we systematically investigate the dynamical evolution of the model in different regimes and uncover the underlying physical mechanisms.


The HTC model extends the TC framework by incorporating vibrational (phonon) degrees of freedom in addition to the strong collective coupling between $N$ two-level excitonic sites (qubits) and a single cavity mode \cite{b57}. The Hamiltonian of the HTC model is composed of three terms,
\begin{equation}\label{eq1}
\hat{H}_{\rm HTC}=\hat{H}_{\rm TC}+\hat{H}_{\rm R}+\hat{H}_{\rm I}.
\end{equation}
The TC Hamiltonian is represented by the expression  ($\hbar$=1)
\begin{equation}
\hat{H}_{\rm TC}= \omega_{c} \hat{a}^{\dagger}\hat{a} + \sum_{n=1}^{N} \left[ \omega_{n}\hat{\sigma}_{n}^{+}\hat{\sigma}_{n}^{-} + g_{n} \left( \hat{a}^{\dagger}\hat{\sigma}_{n}^{-} + \hat{a}\hat{\sigma}_{n}^{+} \right) \right],
\end{equation}
where $\omega_{c}$ is the cavity-mode frequency, $\hat{a}^{\dagger}(\hat{a})$ represents the cavity-mode creation (annihilation) operator, $\omega_{n}$ denotes the transition frequency of the $n$-th qubit, $\hat{\sigma}_{n}^{+}=|1_{n}\rangle\langle 0_{n}|$ $(\hat{\sigma}_{n}^{-}=|0_{n}\rangle\langle 1_{n}|)$ is the raising (lowering) Pauli operator acting on the $n$-th two-level system, and
$g_{n}=\omega_{R}^{n}/\sqrt{N}$ is the corresponding qubit-cavity coupling strength, which reduces to the Rabi frequency $\omega_{R}$ in the
case of a single qubit.
The phonon bath is described by the Hamiltonian
\begin{equation}
\hat{H}_{\rm R}= \sum_{k}\omega_{k}\hat{b}_{k}^{\dagger} \hat{b}_{k}
\end{equation}
and the exciton-phonon interaction is written as
\begin{equation}
\hat{H}_{\rm I}=- \frac{\lambda}{\sqrt{N}} \sum_{k} \sum_{n=1}^{N} \omega_{k} \hat{\sigma}_{n}^{+} \hat{\sigma}_{n}^{-} \left( e^{-i k n} \hat{b}_{k}^{\dagger} + e^{i k n} \hat{b}_{k} \right),
\end{equation}
where $\omega_{k}$ is the frequency of the $k$-th phonon mode, $\hat{b}_{k}^{\dagger}(\hat{b}_{k})$ is the creation (annihilation) operator for the $k$-th phonon mode, and $\lambda$ denotes the diagonal exciton-phonon coupling. In this work, a linear phonon dispersion is adopted,
\begin{equation}
\omega_{k}=\omega_{k_0}\left[ 1+\Gamma \left( \frac{2|k|}{\pi} -1 \right) \right],
\end{equation}
where $\omega_{k_0}$ is the central frequency of the phonon band, $\Gamma$ is a dimensionless bandwidth parameter, and the phonon momentum is given by the expression $k=2 \pi l/N$, $l=-N/2+1,...,N/2$.
To grasp the essential physics of the multi-mode coupling while maintaining analytical tractability, we restrict our analysis to the minimal case of $N=2$, corresponding to $k=0$ and $\pi$.

The standard HTC Hamiltonian described above captures the fundamental interplay among cavity photons, molecular excitons, and vibrational phonons. However, it inherently assumes a perfectly harmonic cavity field and a closed quantum system. To explore rich nonlinear quantum optical phenomena, such as photon blockade and the suppression of multiphoton excitations, and to account for the intrinsic optical nonlinearities of realistic intracavity media, it is essential to introduce an anharmonic energy-level structure for the cavity mode. This is achieved by incorporating a Kerr-type nonlinearity, which physically originates from the third-order optical susceptibility of the medium, thereby extending the standard HTC model to the Kerr-HTC model. Furthermore, realistic molecular cavity platforms are inherently open quantum systems. To probe the nonequilibrium dynamics and sustain steady-state photon emissions, the system must be continuously driven by an external coherent field while simultaneously being subject to cavity photon leakage (dissipation). Taking all these physical ingredients into account, the effective
non-Hermitian Hamiltonian of the driven dissipative Kerr-HTC system can be
expressed as 
\begin{equation}
    \hat{H}_{\text{eff}} = \hat{H}_{\text{HTC}}' + \hat{H}_{\text{Kerr}} + \hat{H}_{\text{drive}}.
    \label{eq:H_eff}
\end{equation}
Here
\begin{equation}\label{Diss}
\hat{H}_{\text{HTC}}' = \hat{H}_{\text{HTC}} - i(\kappa/2)\hat{a}^\dagger\hat{a}
\end{equation}
is the dissipative extension of the standard HTC Hamiltonian
of Eq.~(\ref{eq1}), with $\kappa$ being the cavity decay rate and $1/\kappa$ being the  lifetime of the cavity mode.
This effective non-Hermitian Hamiltonian neglects quantum jumps (fluctuations)  which can be treated, e.g., with  the full Lindblad master equation. However, this approximation  is well justified for the following reasons.
First, quantum jumps  renormalize the steady-state
density matrix, but do not modify the eigenstates
or the energy spectrum of $\hat{H}_{\text{eff}}$, which govern
the essential coherent dynamics.
Second, in the weak dissipation regime considered in this work
($\kappa \ll \omega_R$), the quantum-jump contributions are of higher order in $\kappa$, and their inclusion would
only yield minor quantitative corrections  without altering any of the qualitative phenomena
reported in this paper.

The Kerr nonlinearity Hamiltonian is given by the expression \cite{b58}
\begin{equation}\label{Kerr}
    \hat{H}_{\text{Kerr}} = \chi(\hat{a}^\dagger \hat{a})^2,
\end{equation}
where $\chi$ is the Kerr coefficient. It introduces a photon-number-dependent shift that
breaks the equidistant spacing of the cavity energy levels.
The final term in Eq. (\ref{eq:H_eff}) accounts for the coherent pumping of the cavity
mode by an external laser field \cite{b34},
\begin{equation}\label{drive}
    \hat{H}_{\text{drive}} = \Omega(\hat{a}^\dagger e^{-i\omega_d t} + \hat{a} e^{i\omega_d t}),
\end{equation}
with driving amplitude $\Omega$ and frequency $\omega_d$.

\label{method}

It is essential that the multi-$\rm D_2$  method is applicable to non-Hermitian \cite{b59} and explicitly time-dependent \cite{b60} Hamiltonians.
Hence, high-precision quantum dynamical simulations can be  achieved by applying the time-dependent variational principle with the multi-$\rm D_2$ Ansatz as a trial state for the wavefunction generated by the extended HTC Hamiltonian
of Eq.~(\ref{eq:H_eff}).

The multi-$\rm D_2$ method exhibits systematic convergence: by increasing the multiplicity $M$, the  multi-$\rm D_2$ approaches the numerically accurate solution of the many-body Schr\"odinger equation.
The  Ansatz can be written in the form
\begin{equation}\label{mD2}
	\begin{split}
	&\left| D_{2}^{M}(t) \right\rangle=\\
	&\sum_{n} |n\rangle\sum_{m=1}^{M} A_{mn}(t) \,e^{ \sum_{k}\left(  f_{mk}(t) \, \hat{b}_{k}^{\dagger}-f_{mk}^*(t)\hat{b}_{k} \right) }|0\rangle_{\rm ph},
	\end{split}
\end{equation}
where   $|n\rangle$ are the basis states, $|0\rangle_{\text{ph}}$ denotes the phonon vacuum state,  $A_{mn}(t)$ represents the amplitude of the $n$-th basis state in the $m$-th coherent state component, and $f_{mk}(t)$ is the displacement of the phonon mode with momentum $k$.

The basis states $|n\rangle$ of the  multi-$\rm D_2$ Ansatz of Eq. (\ref{mD2}) are specified according to the following considerations.  
In this work, we focus on the weak-driving regime,  $\Omega\ll\omega_R^{n}$, in which the cavity-mode population is primarily restricted to low-excitation subspaces. Furthermore, the presence of Kerr nonlinearity introduces an anharmonic energy-level structure, which imposes additional energy penalties on multiphoton transitions and further suppresses the population of high-excitation states. Indeed,  the energy of the Kerr-shifted photonic state with \(j\) cavity photons is 
\begin{equation}\label{Ej0}
E_j = \omega_c j + \chi j^2 .
\end{equation}
Hence, the photon of frequency $\omega_d$ is resonant with the transition \(j \to j+1\) if
\begin{equation}\label{Ej}
\omega_d = \omega_c + \chi(2j+1).
\end{equation}
If, for example,  the photon is resonant with the \(0 \to 1\) transition ($j=0$ in Eq. (\ref{Ej})), it has pronounced detunings
\begin{equation}
\delta_j = [\omega_c + \chi(2j+1)] - [\omega_c + \chi]  = 2\chi j
\end{equation}
 with higher-lying transitions (\(j=1, \, 2, ... \)). Hence the higher-lying states are increasingly off-resonant and are strongly blockaded by the Kerr nonlinearity. These states can only  be populated by higher-order multi-photon processes, the contribution of which is negligible in the weak-driving regime considered here. Consequently, we employ the Fock states rather than coherent states to describe the photonic degrees of freedom and assume that the system is confined within  zero-, one-, and two-excitation manifolds. The states in these manifolds can be denoted as $|n_{\text{pt}}, n_{q1}, n_{q2}\rangle$, where $n_{\text{pt}}$ is the cavity photon occupation  number, while $n_{q1}, n_{q2} \in \{0, 1\}$ denote the occupation numbers of the two qubits ($0$ and $1$ correspond to the ground  and excited state, respectively).  With this notation, the zero-excitation  manifold consists of a single vacuum state $|000\rangle$. The one-excitation manifold accommodates three states: $|100\rangle$ (photon is in the state $1_{\text{pt}}$, both qubits are  in the ground state), $|010\rangle$ (photon is in the state $0_{\text{pt}}$, qubit 1 is excited), $|001\rangle$ (photon in the state $0_{\text{pt}}$, qubit 2 is excited). The two-excitation manifold accommodates four states:
$|200\rangle$, $|110\rangle$, $|101\rangle$, and $|011\rangle$.
 These eight states are relabeled as $|n\rangle$ ($n=1,\dots,8$) and yield the basis states of the  multi-$\rm D_2$ Ansatz.

The variational parameters $A_{mn}(t)$ and $f_{mk}(t)$ of $\left| D_{2}^{M}(t) \right\rangle$  in Eq. (\ref{mD2}) are collectively denoted as $\mu_{i}(t)$. Their time evolution is determined by the Dirac-Frenkel time-dependent variational principle
\begin{equation}
\frac{d}{dt} \frac{\partial {L}}{\partial \dot{\mu}_i^*} - \frac{\partial{L}}{\partial \mu_i^*} = 0
\end{equation}
with the Lagrangian
\begin{equation*}
\begin{aligned}
L = i \langle D_2^M(t) | \frac{d}{dt} | D_2^M(t) \rangle - \langle D_2^M(t) | \hat{H}_{\rm eff} | D_2^M(t) \rangle.
\end{aligned}
\end{equation*}

The  equations of motion for the variational parameters $\mu_{i}(t)$ of the driven dissipative Kerr-HTC model are derived in \textit{Supporting Information}. Once $\mu_{i}(t)$ are obtained, the time evolution of the ground-state population can be computed as follows:
\begin{equation}
P_{0}(t)=\sum_{m,m'=1}^{M} A_{m1}^{*}(t) A_{m'1}(t) S_{mm'}(t).
\end{equation}
The population of the one-excitation manifold is
\begin{equation}
P_{1}(t)=\sum_{m,m'=1}^{M}\sum_{n=2}^{4} A_{mn}^{*}(t) A_{m'n}(t) S_{mm'}(t),
\end{equation}
while the population of the two-excitation manifold reads
\begin{equation}
P_{2}(t)=\sum_{m,m'=1}^{M}\sum_{n=5}^{8} A_{mn}^{*}(t) A_{m'n}(t) S_{mm'}(t).
\end{equation}
Finally,
\begin{equation}
\begin{split}
P_{\rm cav}(t)
&=\sum_{m,m'=1}^{M} A_{m2}^{*}(t) A_{m'2}(t) S_{mm'}(t)\\
&+2\sum_{m,m'=1}^{M} A_{m5}^{*}(t) A_{m'5}(t) S_{mm'}(t)\\
&+\sum_{m,m'=1}^{M} A_{m6}^{*}(t) A_{m'6}(t) S_{mm'}(t)\\ &+\sum_{m,m'=1}^{M} A_{m7}^{*}(t) A_{m'7}(t) S_{mm'}(t)
\end{split}
\end{equation}
is the mean cavity photon number, where the coefficient 2 in the second term comes from the two-photon state $|200\rangle$. In the above formulas,
\begin{equation}
S_{mm'} = \exp\sum_k \left\{ -\frac{|f_{mk}|^2 + |f_{m'k}|^2}{2} + f_{mk}^* f_{m'k} \right\}
\end{equation}
is the Debye-Waller factor, a standard object in the DA literature. Physically, it corresponds to the coherent‑state overlap integral.


To ensure clarity and conciseness, we first specify the model parameters used throughout this paper. Unless otherwise stated, we adopt a unified energy scale by setting the cavity frequency $\omega_{c}=1\,\text{eV}$, the driving frequency $\omega_{d}=1\,\text{eV}$, and the qubit frequencies $\omega_{1}=\omega_{2}=1\,\text{eV}$. This corresponds  to the resonant conditions in the ideal Kerr-free ($\chi=0$) system. The qubit-cavity coupling rates are $\omega_{R}^{1}=\omega_{R}^{2}=0.1\,\text{eV}$ ($g_1=g_2=g=0.071\,\text{eV}$, strong-coupling regime), the Kerr coefficient is $\chi=0.1\,\text{eV}$ (moderate Kerr nonlinearity), and the exciton-phonon coupling parameters are set to $\lambda=2$, $\omega_{k_{0}}=0.0124\,\text{eV}$, and $\Gamma=0.3$. The coherent driving strength is relatively weak, $\Omega=0.02\,\text{eV}$.

\begin{figure}[tbph!]
\centering
\includegraphics[scale=0.6,trim=0 0 0 0]{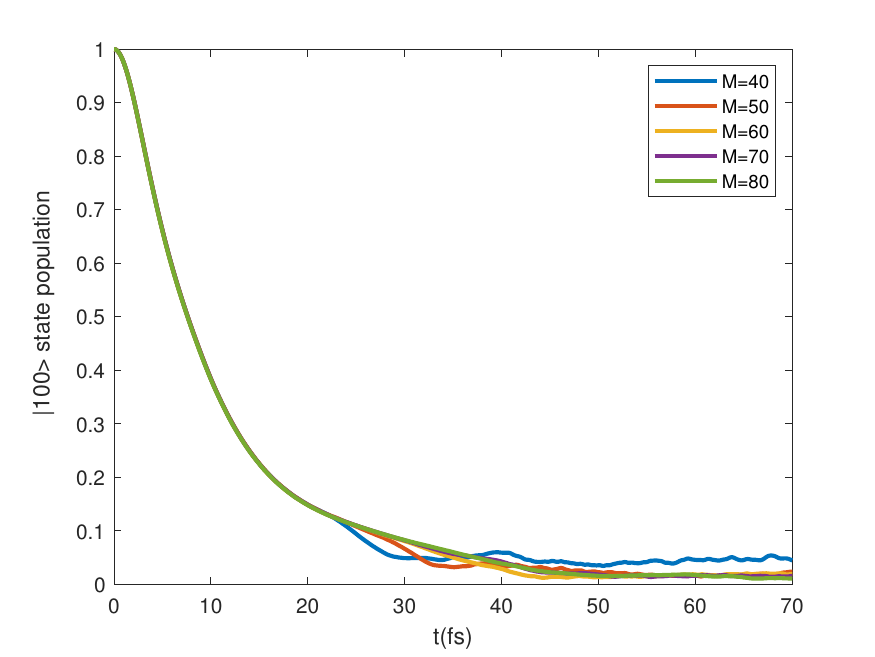}
\caption{Convergence test: The time evolution of the mean cavity photon number $P_{\rm cav}(t)$ for different multiplicities $M$. The system is initialized in $|\psi(0)\rangle=|100\rangle$ under dissipationless conditions ($\kappa=0$). All other parameters are as specified in the main text.}
\label{Fig1}
\end{figure}

Before presenting the main results, we verify the numerical convergence of the multi-$\rm D_{2}$ Ansatz. In the representative dissipation-free case  ($\kappa=0$), we calculate the time evolution of the mean cavity photon number $P_{\rm cav}(t)$ starting from the initial state $|\psi(0)\rangle=|100\rangle$ (i.e., one cavity photon is excited while both qubits are in the ground state) for different multiplicities $M$. As shown in Fig.~\ref{Fig1}, the population dynamics is converged for $M\geq 70$. Therefore, we adopt $M=70$ in all subsequent calculations to ensure both accuracy and computational efficiency.

Before delving into the results of specific simulations, a general comment is appropriate. The cavity driving governed by the Hamiltonian $\hat{H}_{\text{drive}}$ of Eq. (\ref{drive}) breaks conservation of the number of excitations in the  model.
Hence, the driven Kerr-HTC system
exhibits energy exchange between the zero-, one-, and two-excitation manifolds.
This phenomenon is clearly illustrated by the results presented  below. Without driving, the number of excitations  is conserved in the Kerr system \cite{b54,b58}.



\begin{figure}[!htbp]
	\centering
	\includegraphics[scale=0.5]{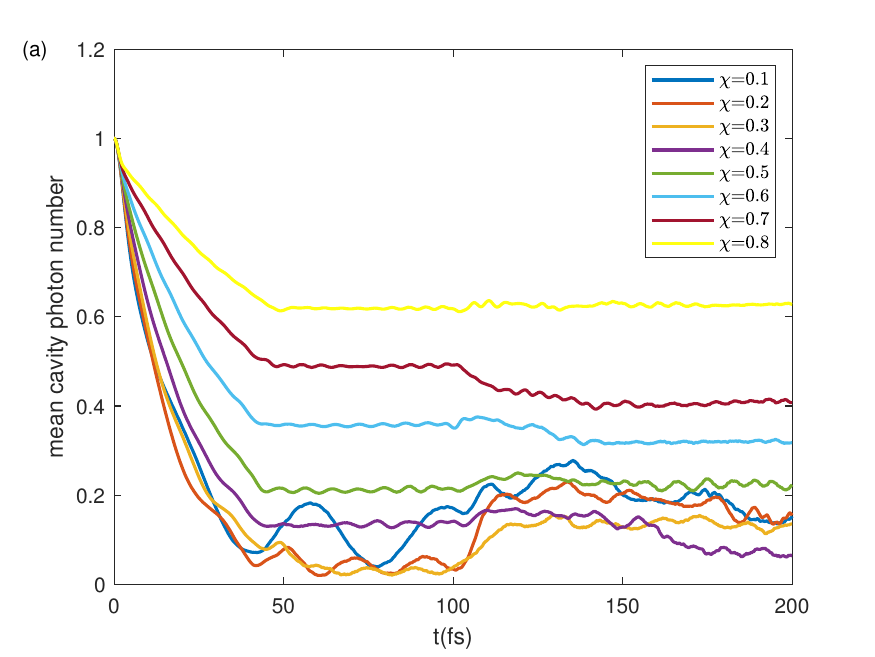}
    \includegraphics[scale=0.5]{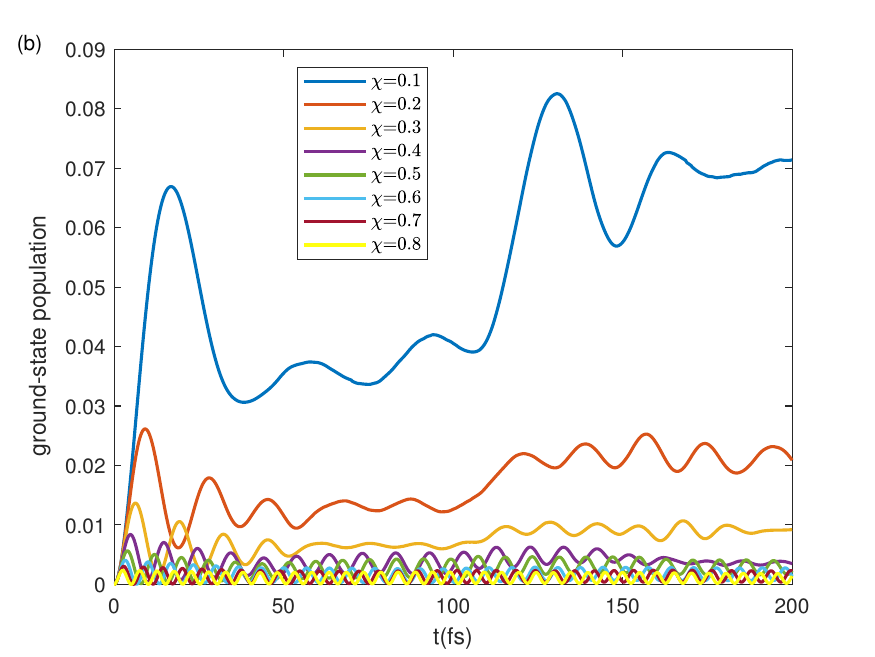}
    \includegraphics[scale=0.5]{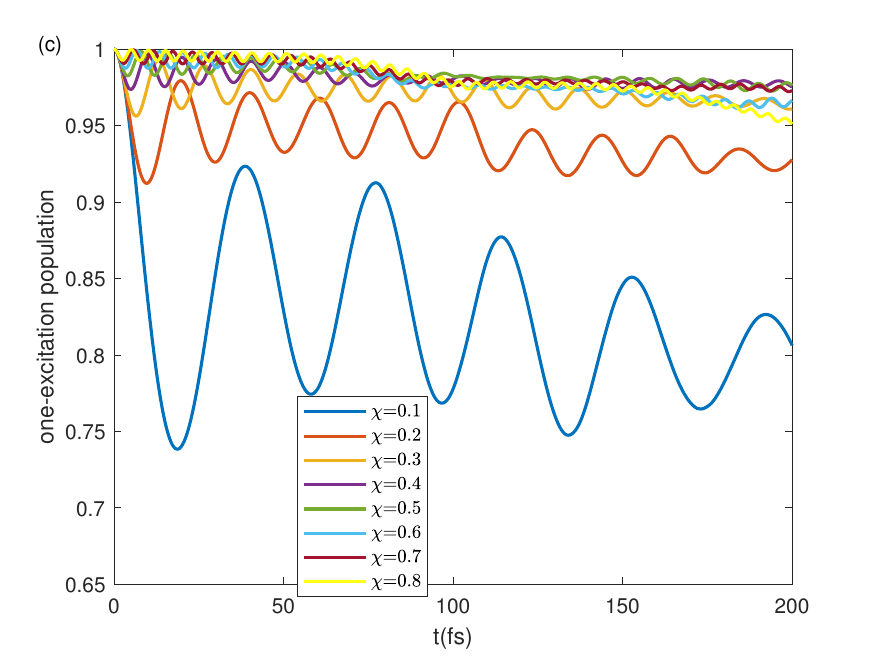}
    \includegraphics[scale=0.5]{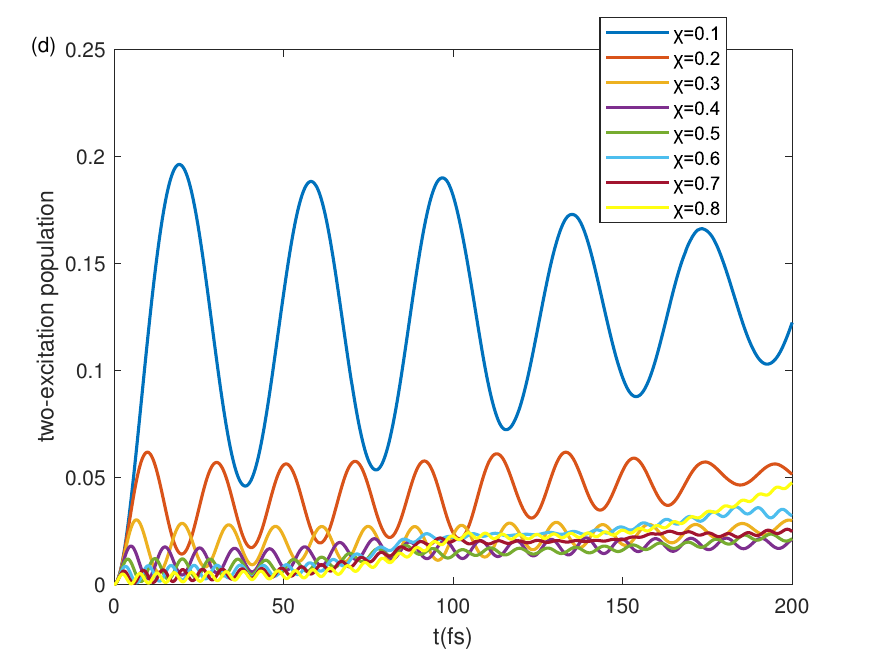}
	\caption{The time evolution of the (a) mean cavity photon number $P_{\rm cav}(t)$, (b) ground-state population $P_{0}(t)$, (c) one-excitation population $P_{1}(t)$, and (d) two-excitation population $P_{2}(t)$ for several Kerr nonlinearities $\chi$ indicated in the panels. Fixed parameters: $\kappa=0$, $\Omega=0.02~\rm eV$, $\omega_d=1~\rm eV$.}
	\label{Fig2}
\end{figure}

To elucidate the role of Kerr nonlinearity, we set $\kappa=0$ (no dissipation) and apply a weak coherent driving with $\Omega=0.02\,\text{eV}$ and $\omega_{d}=1\,\text{eV}$. The system is initialized in the single-photon state $|\psi(0)\rangle=|100\rangle$. We systematically vary the Kerr coefficients $\chi$ from 0.1 eV  to 0.8\,eV and monitor the population dynamics across different excitation manifolds, as shown in Fig.~\ref{Fig2}.

As depicted in Fig.~\ref{Fig2}(a), the mean cavity photon number $P_{\rm cav}(t)$ drops rapidly from its initial value of unity for all Kerr nonlinearities $\chi$, reflecting the fast coherent transfer of the cavity-photon energy into the excitonic degrees of freedom via the light-matter coupling $g$. Following this initial depletion, the photon population exhibits persistent Rabi oscillations for small  Kerr nonlinearities.   As $\chi$ increases, the amplitude of these oscillations drastically decreases. Crucially, the time-averaged photon number increases with $\chi$, indicating that the Kerr nonlinearity acts as an effective barrier against photon leakage into other Fock states, thereby retaining excitation within the cavity mode.

The physical mechanism underlying this behavior is the energy-level anharmonicity induced by the Kerr term $\hat{H}_{\text{Kerr}}$ of Eq. (\ref{Kerr}). For a harmonic cavity ($\chi = 0$), the Fock-state spectrum is equidistant ($E_j = j\omega_c$), allowing resonant multiphoton transitions. The Kerr interaction introduces a photon-number-squared energy shift, Eq. (\ref{Ej0}), which modifies the total energy of the $j$-photon Fock state.
This has two physically distinct consequences. First, the Kerr term assigns a  nonzero self-energy to  the single-photon state, $\langle 100|\hat{H}_{\text{Kerr}}|100\rangle = \chi$. The transition energy from the vacuum $|000\rangle$ state to the single-photon state $|100\rangle$ is therefore shifted from $\omega_c$ to $\omega_c + \chi$ ($j=0$ in Eq. (\ref{Ej})), introducing a Kerr-induced detuning $\chi$ with respect to the driving frequency $\omega_d = \omega_c$. This self-shift partially decouples the single-photon state from the resonant drive, providing a self-protection mechanism that suppresses stimulated photon emission into the vacuum. Second, the two-photon state acquires a four times larger energy shift of $\langle 200|\hat{H}_{\text{Kerr}}|200\rangle = 4\chi$, and the effective transition energy from the single-photon state to the two-photon state becomes  $\omega_c + 3\chi$ ($j=1$ in Eq. (\ref{Ej})).
The anharmonic energy penalty for the second photon thus creates the detuning  $3\chi$. This stronger anharmonicity more effectively breaks the resonance condition for the $|100\rangle \to |200\rangle$ transition, decoupling the two-excitation manifold from the dynamics.

These mechanisms are responsible for the behavior of other observables. The ground-state populations $P_0(t)$ shown in Fig.~\ref{Fig2}(b)  remain small ($< 0.1$) and decrease  with larger $\chi$. Physically, the transition from $|100\rangle$ back to $|000\rangle$ involves an energy change of $\omega_c + \chi$, which is detuned from the driving frequency $\omega_d = \omega_c$ by $\chi$. As $\chi$ grows, this detuning increasingly suppresses the stimulated emission from $|100\rangle$ to $|000\rangle$. Moreover, the cascade pathway $|200\rangle \to |100\rangle \to |000\rangle$ is doubly inhibited: the upper step is blocked by the $3\chi$ anharmonic penalty, and the lower step is off-resonant by $\chi$. Consequently, the system is prevented from returning to the vacuum state. Hence, the ground-state population is substantially depleted, being driven toward zero for large $\chi$.

  Fig.~\ref{Fig2}(c), which shows the population dynamics in the one-excitation manifold, $P_1(t)$, demonstrates that the subspace $\{|100\rangle, |010\rangle, |001\rangle\}$ becomes quasi closed: upward leakage is blocked by the detuning $3\chi$ while downward leakage is off-resonant by $\chi$. Hence  the population oscillates coherently among cavity-photon and excitonic qubits via the coupling $g$, with population approaching unity for large $\chi$.

   Fig.~\ref{Fig2}(d) reveals that the population of the two-excitation manifold, $P_2(t)$, is dramatically suppressed with increasing $\chi$. From a perturbative perspective, the effective coupling between the one- and two-excitation manifolds is proportional to the qubit-photon coupling $g$ and inversely proportional to the anharmonic detuning  $3\chi$. The two-excitation population therefore scales as
  $ P_2(t) \sim [g/(3\chi)]^2$  in the strong nonlinearity regime $\chi \gg g$.

Notably, the population oscillations in Figs.~\ref{Fig2}(c) and (d) are governed by the effective one-excitation Rabi splitting. This behavior can be understood by projecting the dynamics onto the one-excitation subspace \(\{|100\rangle, |010\rangle, |001\rangle\}\), which -- as explained above -- becomes dynamically closed in the strong Kerr-nonlinearity regime. In the rotating frame at \(\omega_d = \omega_c\), the effective Hamiltonian reads
\begin{equation}
H_{\mathrm{eff}}^{(1)}
=
\chi |100\rangle\langle 100|
+
g_{\rm eff}\bigl(|100\rangle\langle 010| + |100\rangle\langle 001| + \mathrm{H.c.}\bigr),
\label{eq:H_eff_1}
\end{equation}
where the diagonal term \(\chi\) arises from the Kerr self-energy, and \(g_{\rm eff}\) is the \emph{renormalized} light-matter coupling. Diagonalization yields two bright eigenstates with the energy splitting
\begin{equation}
\Omega_{\mathrm{eff}}
=
\sqrt{\chi^2 + 8g_{\rm eff}^2},
\label{eq:Omega_eff}
\end{equation}
which governs the coherent population exchange and the corresponding Rabi oscillations. In the limit \(\chi \gg g_{\rm eff}\), Eq.~\eqref{eq:Omega_eff} reduces to \(\Omega_{\mathrm{eff}} \approx \chi\), so the oscillation frequency becomes asymptotically proportional to \(\chi\). Physically, the Kerr self-energy then acts as an effective detuning between the cavity mode and the excitonic qubits, and the system undergoes large-detuning Rabi oscillations whose period is dictated by the nonlinear energy penalty \(\chi\) rather than the coupling \(g_{\rm eff}\).

It is crucial to distinguish the bare coupling \(g\) from the effective coupling \(g_{\rm eff}\). The strong exciton-phonon coupling \((\lambda = 2)\) inherent in the present model induces polaron formation: the excitonic states are dressed by phonon clouds, and the light-matter coupling is reduced,
\begin{equation}
g_{\rm eff}
=
g\,e^{-\lambda^2/2}.
\label{eq:g_eff}
\end{equation}
For the present parameters, $g_{\rm eff} \approx 0.0096$ eV,
which is  much smaller than the Kerr coefficient \(\chi = 0.1\,\mathrm{eV}\). Substituting \(g_{\rm eff}\) into Eq.~\eqref{eq:Omega_eff} yields $\Omega_{\mathrm{eff}} = 0.104$ eV,
corresponding to a period \(T = 2\pi/\Omega_{\mathrm{eff}} \approx 40\,\mathrm{fs}\). This is in excellent agreement with the population oscillations observed in Fig.~\ref{Fig2}(c) for \(\chi = 0.1\,\mathrm{eV}\). It also demonstrates that the oscillation period of the blue curve in Fig.~\ref{Fig2}(c) is not governed by the bare coupling \(g\), which would give \(18.4\,\mathrm{fs}\), but rather by the phonon-renormalized coupling \(g_{\rm eff}\), which makes \(\Omega_{\mathrm{eff}} \approx \chi\) and yields \(40\,\mathrm{fs}\).

As \(\chi\) increases, the condition \(\chi \gg g_{\rm eff}\) remains well satisfied, and Eq.~\eqref{eq:Omega_eff} becomes increasingly accurate. For example, at \(\chi = 0.8\,\mathrm{eV}\), Eq.~\eqref{eq:Omega_eff} gives \(\Omega_{\mathrm{eff}} \approx 0.825\,\mathrm{eV}\), corresponding to a period of about \(5\,\mathrm{fs}\), in agreement with the fast oscillations observed in Fig.~\ref{Fig2}(c). The identical oscillation frequency observed in the two-excitation manifold [Fig.~\ref{Fig2}(d)] arises because its dynamics are ``slaved'' to the one-excitation subspace: \(P_2(t)\) is parametrically driven by \(P_1(t)\) and inherits the same characteristic frequency.



\begin{figure}[!htbp]
	\centering
	\includegraphics[scale=0.5]{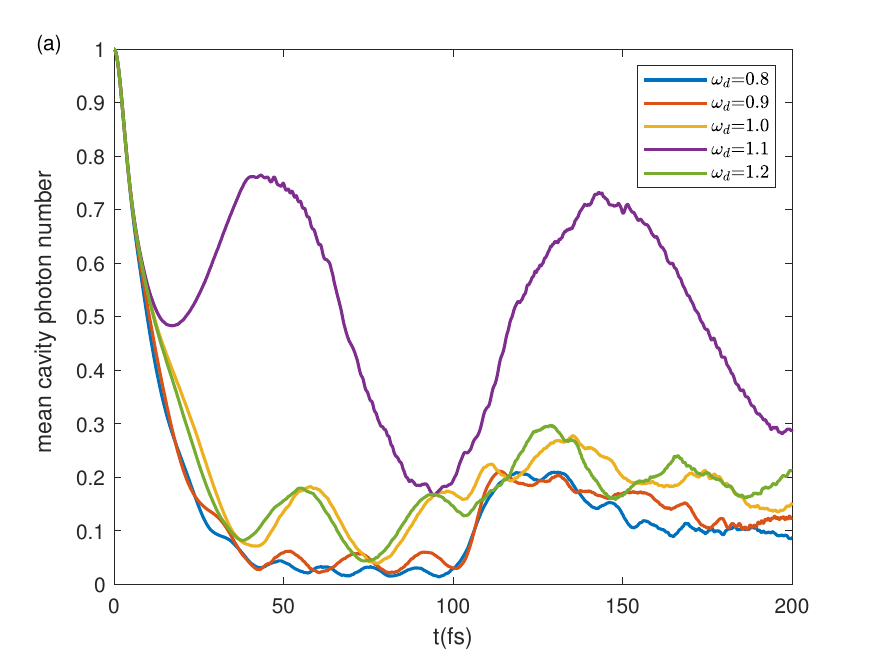}
	\includegraphics[scale=0.5]{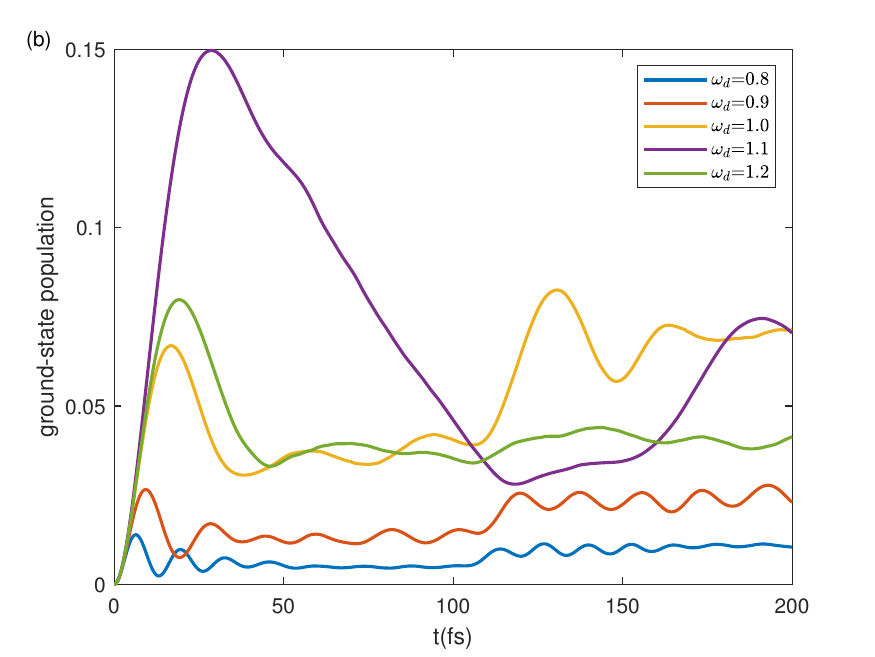}
	\includegraphics[scale=0.5]{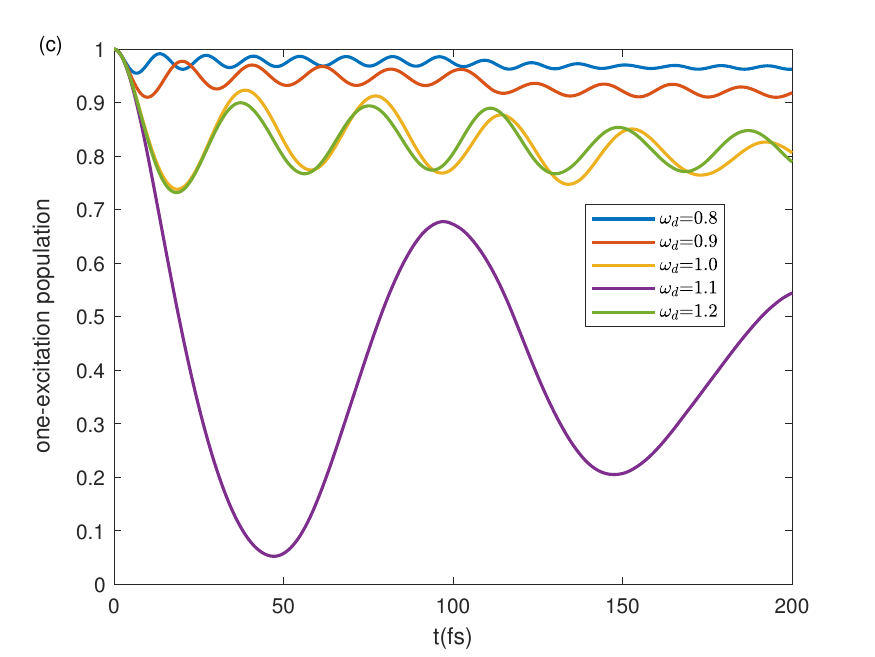}
	\includegraphics[scale=0.5]{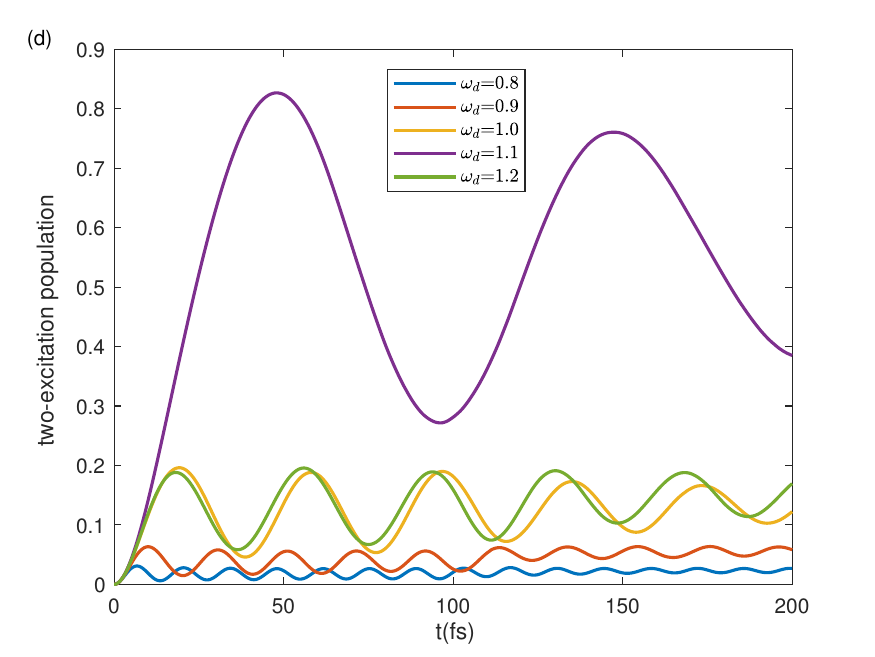}
	\caption{The time evolution of the (a) mean cavity photon number $P_{\rm cav}(t)$, (b) ground-state population $P_{0}(t)$, (c) one-excitation population $P_{1}(t)$, and (d) two-excitation population $P_{2}(t)$ for various driving frequencies $\omega_d$, as indicated. The fixed parameters are $\kappa=0$, $\Omega=0.02\,\text{eV}$, and $\chi=0.1\,\text{eV}$.}
	\label{Fig3}
\end{figure}

To investigate the effect of the driving frequency on the system dynamics, we set $\kappa = 0$, $\Omega = 0.02\,\text{eV}$, and $\chi = 0.1\,\text{eV}$. We then examine the time evolution of the mean cavity photon number and the populations of the ground, one-excitation, and two-excitation states for driving frequencies $\omega_d = 0.8, 0.9, 1.0, 1.1,$ and $1.2\,\text{eV}$. The results are presented in Fig.~\ref{Fig3}.

For $\omega_d = 0.8$ and $0.9\,\text{eV}$,  the mean cavity photon number remains low after an initial decay (Fig.~\ref{Fig3}(a)). Correspondingly, as seen in Figs.~\ref{Fig3}(b)--(d), the one-excitation population dominates throughout, while the ground-state and the two-excitation populations remain  minimal. This indicates that off-resonant driving at lower frequencies fails to inject energy effectively, confining the system primarily within the one-excitation subspace. As $\omega_d$ increases to $1.0\,\text{eV}$, the system response intensifies, exhibiting larger-amplitude  Rabi  oscillations in $P_{\rm cav}(t)$ and a moderate rise in the ground- and two-excitation populations. This signifies that the  driving begins to facilitate inter-manifold population transfer at the bare resonance conditions.

At $\omega_d = 1.1\,\text{eV}$, the system exhibits its strongest response, characterized by large-amplitude oscillations of $P_{\rm cav}(t)$, a marked rise in the ground- and two-excitation populations, and a substantial oscillatory depletion of the one-excitation population. This identifies $\omega_d = 1.1\,\text{eV}$ as the frequency matching the system's effective energy gap $\omega_{c}+\chi$ between the zero- and one-excitation manifolds. The pure two-photon state \(|200\rangle\) and higher-excitation states such as \(|300\rangle\), \(|210\rangle\), etc., are strongly blockaded by the Kerr nonlinearity. However, it should be noted that the three-excitation manifold also contains single-photon plus two-exciton states, e.g., \(|111\rangle\). The energy difference for absorbing a photon from \(|011\rangle\) to \(|111\rangle\) is still \(\omega_c + \chi = 1.1\,\mathrm{eV}\), which is formally resonant. Nevertheless, reaching \(|011\rangle\) from the initial state \(|100\rangle\) requires multiple driving steps, and is  a higher-order in \(g\) process whose contribution is tiny in the weak-driving regime. Therefore, truncating to the zero-, one-, and two-excitation manifolds is justified in the weak-driving, short-time, single-photon-blockade framework considered here.

At $\omega_d = 1.2\,\text{eV}$, both $P_{\rm cav}(t)$ and the two-excitation population decrease, while the one-excitation population regains dominance, indicating that the system moves away from the effective resonance regime for higher driving frequencies. Note that all the observables shown in Figs.~\ref{Fig3}(a)--(d) for $\omega_d = 1 \,\text{eV}$ and $\omega_d = 1.2 \,\text{eV}$ show remarkable similarity, which reveals the approximate symmetry of the results with respect to the frequency $\omega_d = 1.1\,\text{eV}$ which is  resonant to the effective energy gap.

These results establish the driving frequency $\omega_d$ as a potent control parameter for the system dynamics. The response follows a non-monotonic trend, peaking at the Kerr-modified frequency $\omega_d = 1.1\,\text{eV}$ rather than at the bare frequency of $1.0\,\,\text{eV}$. Hence, the combined interplay of Kerr nonlinearity, cavity-exciton coupling, and exciton-phonon coupling leads to a renormalization of the system's energy level structure, shifting the effective resonance away from the bare cavity frequency.



\begin{figure}[!htbp]
	\centering
	\includegraphics[scale=0.5]{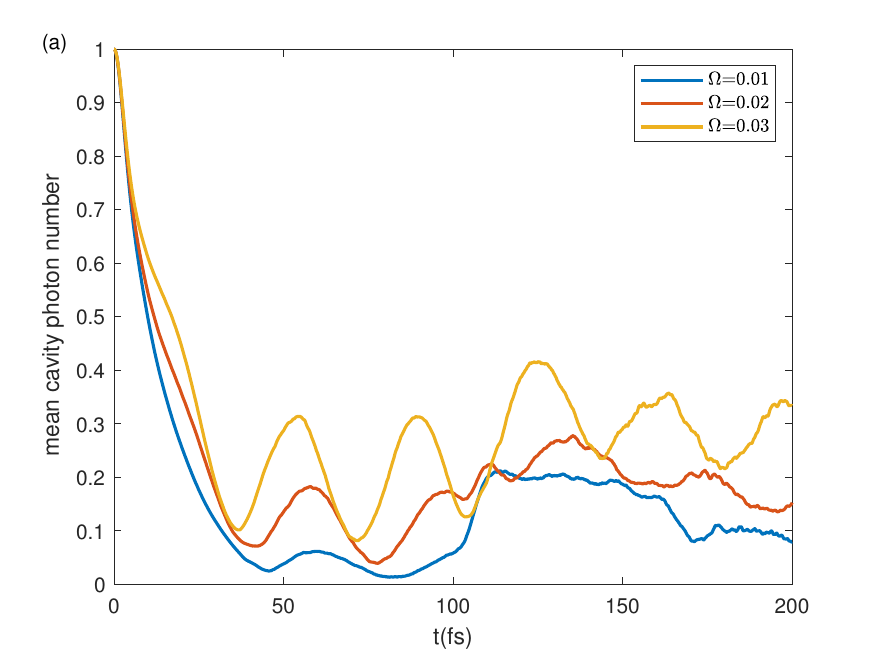}
	\includegraphics[scale=0.5]{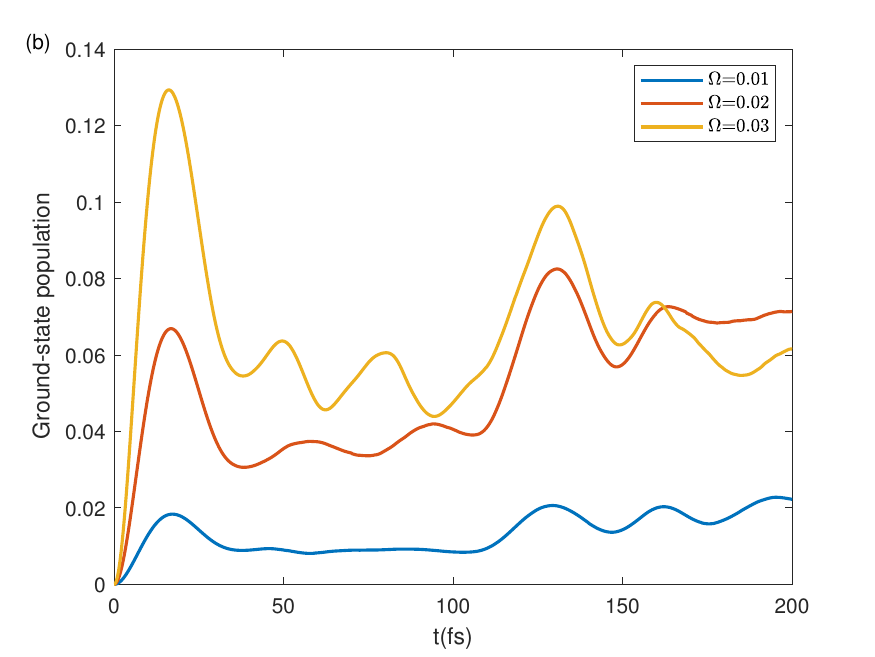}
	\includegraphics[scale=0.5]{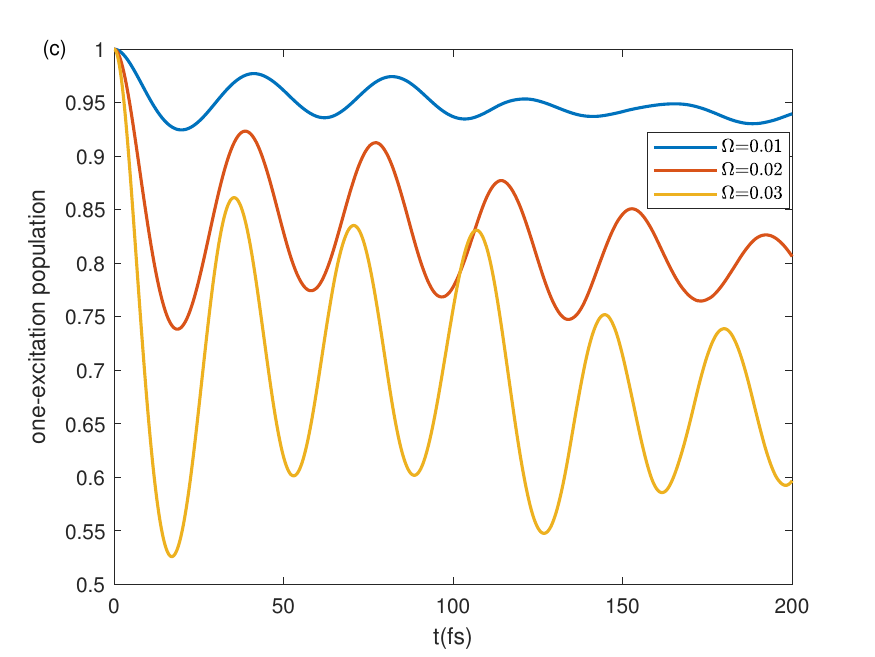}
	\includegraphics[scale=0.5]{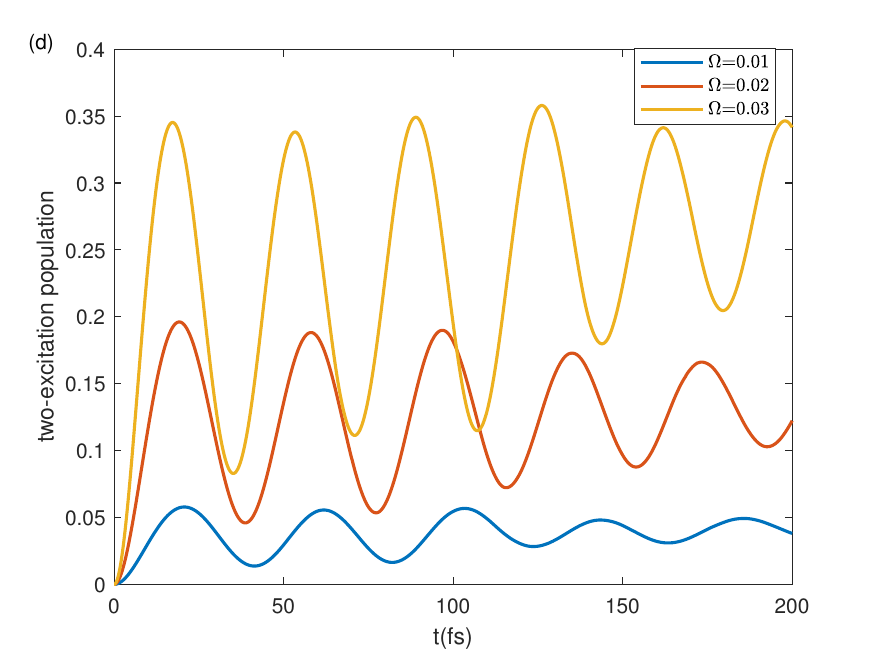}
	\caption{The time evolution of the (a) mean cavity photon number $P_{\rm cav}(t)$, (b) ground-state population $P_{0}(t)$, (c) one-excitation population $P_{1}(t)$, and (d) two-excitation population $P_{2}(t)$ for several  driving strengths $\Omega$. The fixed parameters are $\kappa = 0$, $\omega_d = 1.0\,\,\text{eV}$, and $\chi = 0.1\,\,\text{eV}$.}
	\label{Fig4}
\end{figure}

We now turn to the role of the driving strength $\Omega$. The driving field acts as the external energy source, and its amplitude determines whether the system can overcome the nonlinear barriers to access higher excitation manifolds. To isolate this effect, we set $\kappa = 0$, $\omega_d = 1.0\,\text{eV}$, $\chi = 0.1\,\text{eV}$, and systematically vary $\Omega$ from $0.01$ to $0.03\,\text{eV}$. The resulting dynamics of the mean cavity photon number and the population distributions are presented in Fig.~\ref{Fig4}.

As shown in Fig.~\ref{Fig4}(a), increasing $\Omega$ significantly enhances  the overall magnitude and oscillation amplitude of the mean cavity photon number $P_{\rm cav}(t)$. This  observation is microscopically rooted in the pronounced redistribution of energy across different excitation manifolds, as detailed in Figs.~\ref{Fig4}(b)--(d).

In the weak-driving regime ($\Omega = 0.01\,\text{eV}$), the one-excitation population $P_1(t)$ remains dominant, oscillating robustly around $0.95$ [Fig.~\ref{Fig4}(c)], while the ground-state ($P_0(t)$) and two-excitation ($P_2(t)$) populations remain minimal [Figs.~\ref{Fig4}(b) and (d)]. This behavior is a direct dynamical manifestation of the photon blockade effect discussed above. Physically, the weak driving field is unable to overcome the anharmonic energy penalty $3\chi$ required to populate the two-photon state $|200\rangle$ from the $|100\rangle$ state. Consequently, the one-excitation subspace spanned by $\{|100\rangle, |010\rangle, |001\rangle\}$ acts as a quasi-closed trapping manifold, and the dynamics are essentially restricted to single-photon processes.

As the driving strength increases to $\Omega = 0.02\,\text{eV}$ and $0.03\,\text{eV}$, a qualitative transition occurs. The oscillations in $P_1(t)$ increase, while their amplitudes  and the  time-averaged value of $P_1(t)$ decrease noticeably [Fig.~\ref{Fig4}(c)]. This is  accompanied by a marked rise in both $P_0(t)$ and $P_2(t)$ [Figs.~\ref{Fig4}(b) and (d)]. Notably, at $\Omega = 0.03\,\text{eV}$, the ground-state population $P_0(t)$ exhibits prominent transient maxima reaching approximately $0.12$, and the two-excitation population $P_2(t)$ is also substantially enhanced, exceeding 0.35.

Mechanistically, this transition is governed by the competition between the coherent driving strength \(\Omega\) and the Kerr-induced anharmonicity \(\chi\). The driving Hamiltonian \(\hat{H}_{\text{drive}}\) of Eq. (\ref{drive}) directly couples adjacent photon-number states with a matrix element proportional to \(\Omega\sqrt{j+1}\). For \(\omega_d = 1.0\,\mathrm{eV}\), the near-resonant channels \(|010\rangle \to |110\rangle\) and \(|001\rangle \to |101\rangle\) are detuned only by \(\chi = 0.1\,\mathrm{eV}\). Because of this scaling, a moderate increase of \(\Omega\) to \(0.03\,\mathrm{eV}\) induces power broadening and multi-step virtual processes that efficiently populate these single-photon-exciton states. Consequently, the total two-excitation population \(P_2(t)\) increases significantly. However, the pure two-photon state \(|200\rangle\) remains detuned by \(3\chi = 0.3\,\mathrm{eV}\), so the Kerr blockade of double-photon excitation is not destroyed. As \(\Omega\) becomes comparable to the nonlinear detuning \(\chi\), the driving-induced power broadening bridges the residual detuning of these near-resonant channels, partially relaxing the one-excitation confinement and opening inter-manifold transition channels. Concurrently, the enhanced population in the two-excitation manifold accelerates the cascade processes and cavity-exciton coherent exchange, which in turn feeds population back into the ground state.

Fig.~\ref{Fig4} demonstrates that the driving strength $\Omega$ serves as a crucial control knob for regulating inter-manifold population transfer. By tuning $\Omega$, the system dynamics can be continuously driven from a Kerr-dominated, one-excitation confined regime to a multi-channel coherent evolution regime involving the ground, one-excitation, and two-excitation manifolds. This highlights the delicate balance between energy injection (driving) and energy-level anharmonicity (Kerr nonlinearity) in shaping the cavity system dynamics.



\begin{figure}[!htbp]
	\centering
	\includegraphics[scale=0.5]{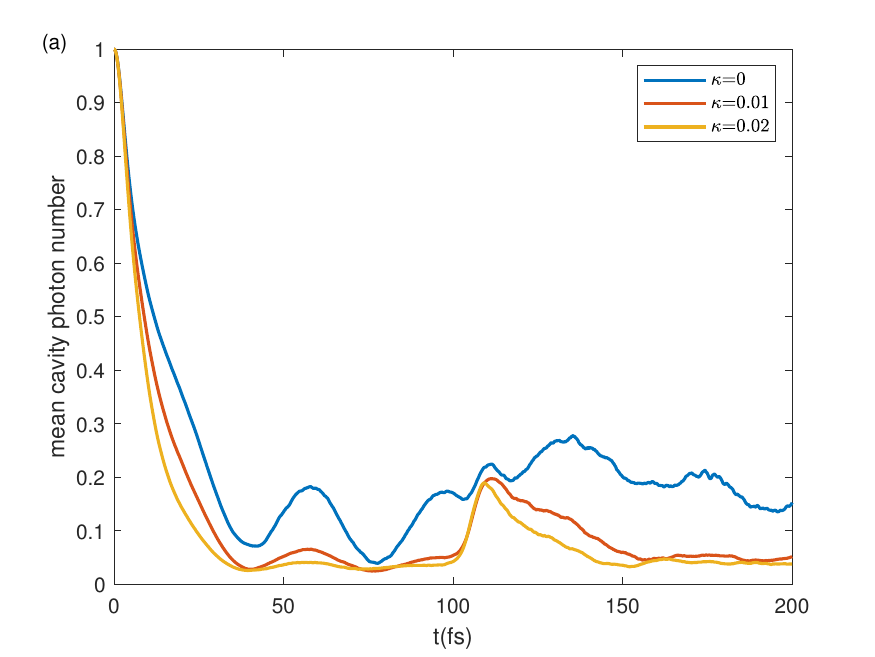}
	\includegraphics[scale=0.5]{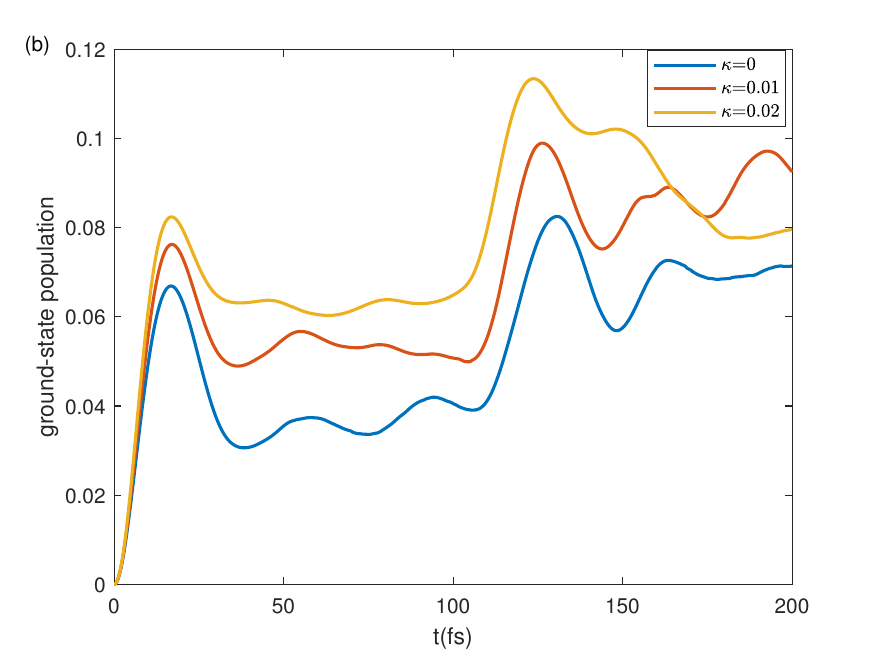}
	\includegraphics[scale=0.5]{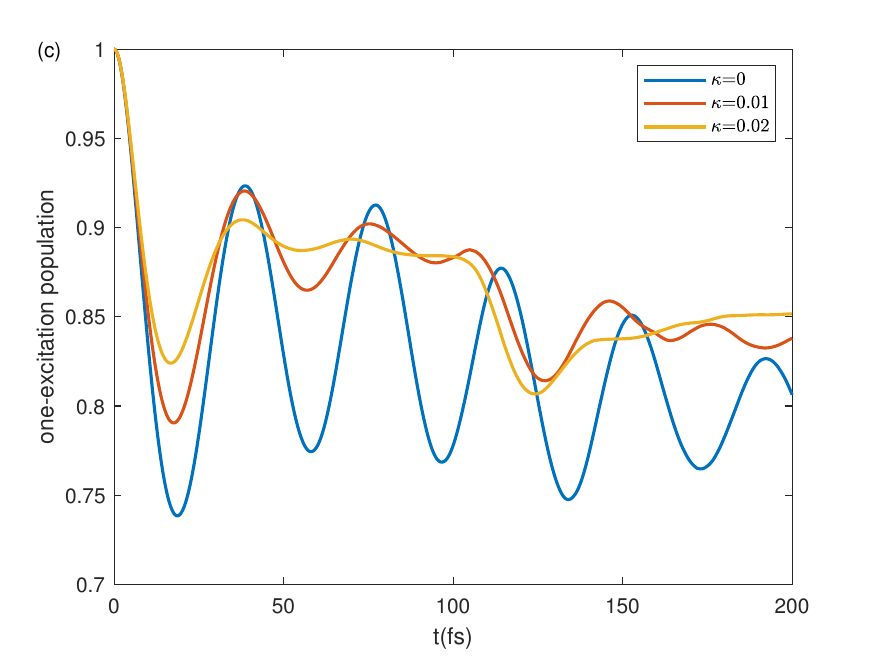}
	\includegraphics[scale=0.5]{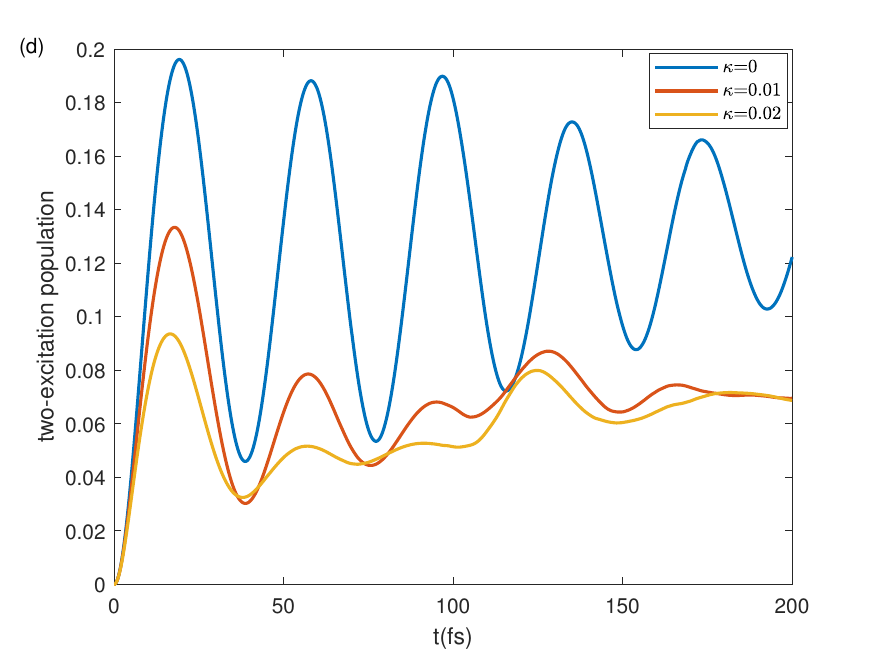}
	\caption{Impact of the cavity decay rate $\kappa$ on the system dynamics. (a) mean cavity photon number $P_{\rm cav}(t)$, (b) ground-state population $P_0(t)$, (c) one-excitation population $P_1(t)$, and (d) two-excitation population $P_2(t)$. The fixed parameters: $\Omega = 0.02\,\text{eV}$, $\omega_d = 1.0\,\text{eV}$, $\chi = 0.1\,\text{eV}$.}
	\label{Fig5}
\end{figure}

In the preceding subsections, we have elucidated the coherent dynamics governed by the interplay of Kerr nonlinearity and external driving in a closed system ($\kappa = 0$). However, realistic molecular cavity platforms are inherently open quantum systems. Therefore, we now introduce the cavity mode decay rate $\kappa$, which acts as an irreversible energy sink. Fixing $\Omega = 0.02\,\text{eV}$, $\omega_d = 1.0\,\text{eV}$, and $\chi = 0.1\,\text{eV}$, we systematically vary $\kappa$ to investigate its impact on the nonequilibrium population transfer, as illustrated by Fig.~\ref{Fig5}.

As shown in Fig.~\ref{Fig5}(a), increasing $\kappa$ accelerates the decrease  of the mean cavity photon number $P_{\rm cav}(t)$, progressively suppresses its oscillation amplitude, and makes the depopulation irreversible. This  damping indicates that cavity dissipation continuously extracts photons from the system, shortening the lifetime of photon-carrying states and weakening the coherent light-matter exchange processes. While the overall photon number decreases, the dissipation exerts a profoundly asymmetric impact on the populations of different excitation manifolds, as detailed in Figs.~\ref{Fig5}(b)--(d).

The irreversible nature of the cavity-mode dissipation dictates that the extracted energy must ultimately be deposited somewhere. As shown in Fig.~\ref{Fig5}(b), the ground-state population $P_0(t)$ increases monotonically with rising $\kappa$. Physically, the irreversible photon leakage cascades the system down the energy ladder. Since the vacuum state $|000\rangle$ contains no cavity photons, it is completely immune to the dissipation operator $-i(\kappa/2)\hat{a}^\dagger\hat{a}$. Consequently, $P_0(t)$ acts as an absorbing boundary, continuously accumulating the population that is drained from the higher-lying, photon-carrying manifolds.

Having established where the lost population ends up, we turn to the states that actively participate in the dissipative dynamics. The one-excitation manifold [Fig.~\ref{Fig5}(c)] demonstrates remarkable robustness. While its oscillation amplitude diminishes markedly with increasing $\kappa$, the total population $P_1(t)$ remains dominant and avoids collapse. This resilience is microscopically rooted in the composition of the one-excitation manifold $\{|100\rangle, |010\rangle, |001\rangle\}$. Although the cavity-photon state $|100\rangle$ is directly subject to the cavity decay rate $\kappa$, the purely excitonic states ($|010\rangle$ and $|001\rangle$) are not directly affected by the cavity-mode dissipation. The rapid coherent exchange mediated by the light-matter coupling effectively shelters a significant fraction of the population in these dissipation-free excitonic states, rendering the one-excitation manifold relatively immune to cavity loss.

In contrast to the resilience of the one-excitation manifold, the two-excitation population [Fig.~\ref{Fig5}(d)] is strongly affected by dissipation. Even at moderate $\kappa$, $P_2(t)$ experiences a drastic suppression, which is caused by two synergistic mechanisms. First, the two-photon state $|200\rangle$ contains two cavity photons, making its effective cavity decay rate proportional to $2\kappa$, which significantly accelerates its depletion compared to single-photon states. Second, recalling our analysis of how the driving strength affects the system dynamics, populating the two-excitation manifold requires the driving field to overcome the Kerr-induced anharmonic penalty of $3\chi$. The presence of dissipation introduces an additional homogeneous energy broadening that further disrupts the resonance condition required for multiphoton transitions. Thus, the two-excitation manifold is squeezed by the dual constraints of nonlinear energy penalties and enhanced dissipative losses.

The contrasting behaviors across Figs.~\ref{Fig5}(b)--(d) collectively reveal that cavity dissipation $\kappa$ functions as far more than a simple decay channel; it acts as a highly selective manifold filter. By preferentially depleting the fragile multi-photon states while leaving the one-excitation manifold structurally intact, dissipation synergizes with the Kerr nonlinearity to actively reinforce the photon blockade effect. Consequently, even when a strong driving field attempts to open multiphoton channels, moderate cavity dissipation can counteract this tendency and restore the system to a robust single-photon-dominated regime. This cooperative mechanism between nonlinearity and dissipation is important for engineering robust single-photon sources and nonclassical light fields in open molecular cavity devices.


In summary, we have established an extended driven-dissipative HTC model incorporating Kerr nonlinearity and employed the multi-$\rm D_2$ method  to systematically investigate the ensuing  cavity-driven dynamics. Rather than treating nonlinearity, driving, and dissipation as isolated processes, our work reveals a fundamental physical feedback loop governed by many-body energy level renormalization, coherent energy injection, and selective dissipative filtering.

Our results demonstrate that the Kerr nonlinearity, in synergy with exciton-phonon and exciton-photon couplings, induces a profound photon-number-dependent anharmonic level structure. This many-body renormalization not only shifts the effective resonance frequency significantly away from the bare cavity mode but also imposes a stringent anharmonic energy penalties  that isolate the one-excitation manifold from the zero- and two-excitation manifolds, respectively -- establishing the foundation for the photon blockade effect.

Building upon this renormalized level structure, the external coherent driving and cavity dissipation play competing yet cooperative roles in shaping the nonequilibrium population transfer. The driving field acts as an energy injection valve: while weak driving respects the nonlinear barriers and maintains single-photon dominance, strong driving induces power broadening that overcomes the anharmonic penalty, opening multiphoton transition channels and breaking the one-excitation confinement. Crucially, the introduction of cavity dissipation counteracts this breakdown. Dissipation functions not only as a  decay channel, but also  as a highly selective manifold filter. By preferentially depleting the fragile multi-photon states (which suffer from enhanced cavity decay rates and disrupted resonance conditions) while sheltering the robust one-excitation population within dissipation-free excitonic dark states, cavity dissipation actively synergizes with the Kerr nonlinearity to reinforce the photon blockade effect.

Ultimately, this cooperative mechanism demonstrates that moderate cavity-mode dissipation can be harnessed to restore and stabilize single-photon-dominated regimes, even in the presence of strong driving fields that attempt to induce multiphoton excitations. These findings not only clarify the intricate interplay among nonlinearity, driving, and dissipation in polariton dynamics but also provide a crucial theoretical blueprint for engineering robust single-photon sources and nonclassical light fields in open molecular cavity devices.

Future work will extend this framework to larger collectively coupled molecular ensembles, incorporate the full Lindblad master equation to account for quantum jump processes in the cavity, as well as calculate the second-order photonic correlation function  and spectral responses to comprehensively characterize the quantum statistical features of the emitted nonclassical light fields.

\section*{Associated Content}
\twocolumngrid
\vspace{-0.45em}
\noindent
\begin{minipage}{\columnwidth}
	\textbf{Data Availability Statement}\par
	\vspace{0.08em}
The data that support the findings of this study are available from the corresponding author upon reasonable request.
\end{minipage}

\vspace{0.40em}
\noindent\textbf{Supporting Information}\par
\vspace{0.10em}
Supporting Information contains kernel formulas.

\vspace{0.40em}
\noindent\textbf{Acknowledgments}\par
\vspace{0.10em}
K. S. acknowledges support from the National Natural Science Foundation of China (Grant No. 22573025). M. F. G. acknowledges support from the National Natural Science Foundation of China (Grant No. 22373028). The authors gratefully acknowledge the support of the Singapore Ministry of Education Academic Research Fund (Grant No. RG2/24).


\twocolumngrid


\end{document}